# A Computational Approach for Designing Tiger Corridors in India

Saurabh Shanu[1*], Sudeepto Bhattacharya[2]

[1]Department of Virtualization, School of Computer Science and Engineering, University of Petroleum and Energy Studies, Dehradun 248007, Uttarakhand, India

[2]Department of Mathematics, School of Natural Sciences, Shiv Nadar University, P.O. Shiv Nadar University, Greater Noida, Gautam Buddha Nagar 201 314, Uttar Pradesh, India

sshanu@ddn.upes.ac.in

**Abstract:** Wildlife corridors are components of landscapes, which facilitate the movement of organisms and processes between intact habitat areas, and thus provide connectivity between the habitats within the landscapes. Corridors are thus regions within a given landscape that connect fragmented habitat patches within the landscape. The major concern of designing corridors as a conservation strategy is primarily to counter, and to the extent possible, mitigate the effects of habitat fragmentation and loss on the biodiversity of the landscape, as well as support continuance of land use for essential local and global economic activities in the region of reference.

In this paper, we use game theory, graph theory, membership functions and chain code algorithm to model and design a set of wildlife corridors with tiger (*Panthera tigris tigris*) as the focal species. We identify the parameters which would affect the tiger population in a landscape complex and using the presence of these identified parameters construct a graph using the habitat patches supporting tiger presence in the landscape complex as vertices and the possible paths between them as edges. The passage of tigers through the possible paths have been modelled as an Assurance game, with tigers as an individual player. The game is played recursively as the tiger passes through each grid considered for the model. The iteration causes the tiger to choose the most suitable path signifying the emergence of adaptability.

As a formal explanation of the game, we model this interaction of tiger with the parameters as deterministic finite automata, whose transition function is obtained by the game payoff.

## 1. Introduction

Landscape linkage may be defined as the extent to which the landscape facilitates movement among resource habitat patches [41, 42]. We also define corridor as a patch, usually linear, embedded in a matrix within a landscape, that connects two or more bigger patches of habitat, thereby providing linkage between the habitats and that is proposed for conservation on the basis that it will improve or maintain the viability of specific wildlife populations in the concerned habitat patches. Further, we define a strategy for selection of passage as travel via a corridor by individuals of the focal species from one habitat patch to another [6].

Wildlife corridors, as implied from the definition above, are integral components of ecological landscapes. The objective of wildlife corridors is to facilitate the movement of processes and organisms between considered areas in the landscape. Corridors are thus regions within a given landscape that generally comprise native vegetation, and connect otherwise disconnected, fragmented, non-contiguous wildlife habitat patches in the focal landscape [6, 11].

Corridors, being integral components of landscapes, are characterized by two distinct categories of components, namely, pattern and process components [11]. The structural corridor and the functional corridors present the categories of wildlife corridors. The structural categorization refers to the geographical existence of the landscape between the focal patches and the functional corridor is a resultant of both – species and landscape. Hence, a functional wildlife corridor refers to both, species - as well as landscape-specific concept. Corridors thus, may be considered as evolving phenomena, caused by the interaction between process and pattern attributes of the area. The essential function and utility of wildlife corridors is thus to connect at least two key habitat areas of biological significance, and thus ensure gene flow between spatially separate populations of species, fragmented due to landscape modifications, by supporting the movements of both biotic and abiotic processes [5, 6, 11, 13].

Researchers have demonstrated that presence of species-specific wildlife corridors within a given landscape to be instrumental in increasing gene flow and population sizes of the species [6, 17, 18, 19, 20].

The above discussions imply that any realistic modelling to design wildlife corridor has to be a species – specific task, with a proper selection of habitat for the concerned focal species. In the present paper, we present a computational procedure for designing corridor for the Indian tiger (*Panthera tigris tigris*) in the Indian landscape. For a country biogeographically as vast and diverse as India, relative spatial location of national reserves with reference to each other becomes an important factor to consider for taking optimal decision on resource allocations, and thus either protecting existing tiger corridors, or even in some instances, creating proper wildlife corridors in. An important objective in such a decision making therefore must be to select the critical tiger habitats (CTH) in a way that their spatial configuration guarantees a high degree of interconnectivity within the intensely human-dominated landscapes, over a long term land use scenario.

One means to achieve the above objective would be to design the interconnectivity among the existing (or even potential) habitats or CTH using a network model. In such a network, each tiger habitat would be treated as a vertex, and the tiger corridors between these vertices would be the edges.

The major purpose of this paper is to provide a basic computational framework for understanding a viable corridor network design within the focal landscape complex for tigers. In this work, all arguments and observations are based on the basic structural definition of connectivity, where the existence and viability of a corridor is required to be understood and determined entirely by the landscape features and structure.

We describe the problem of tiger corridor planning and designing within the landscape as a connection subgraph problem [6]. We next incorporate the conflict of interest between the traversing tiger and the landscape features resultant of primarily anthropogenic modifications, through an Assurance game. Finally, informed about the possible costs, we provide an optimized path and thus use these optimized paths to design a Deterministic Finite Automata to obtain the grammar for designing corridors, which we claim, could serve as a rule base for corridor design.

Although the present work makes reference to a landscape map of the focal complex, it is essentially semi-emperical and schematic in nature. Accordingly, the discussions that follow do not refer or connect to any real-world datasets as would have been obtained through a proper resolution GIS routine. Since the work focuses on the presence or absence of corridors linking various tiger habitats in the complex, the distances involved, and the ease of movement for the tiger through these corridors, we are, however, of the opinion that the work could serve as a schema for an informed decision-making by conservationists and wildlife managers in designing real-world corridors.

Section 2 contains the essentials of the mathematical concepts that have been used in this paper. Sections 3 and 4 describe the modelling and the conclusion of the work, respectively.

## 2. Background for Modelling

In the present work, we shall describe a modelling of a feasible wildlife corridor for the tiger using few specific concerned areas of computational frameworks. In this section, we shall provide the essentials of these areas, in order to make the work self-contained.

We apply game theory to model the effect of presence or absence of identified parameters in a grid leading to selection for movement by tigers. The choice of tigers for movement happens to be random but computationally what must be preferable according to the behavioral pattern of tigers has been modelled here, which could act as an active strategy for designing the corridors. The results are in turn the implications of essentially non-linear computational interactions among the parameters and the focal species.

Assurance game, best represents the present scenario. While modelling the present interactions, we suppose that set of parameters and the focal species are players in the game, and thus accrue a series of pay-off depending on the co-player's as well as its own strategies. The game is recursively repeated over discrete time-steps to produce the complex dependencies of parameters affecting the focal species.

Further, in Assurance game, always a minimal cost is contributed by all the interacting players if they are to obtain any benefit from their own chosen strategy. Thus, such a game would best capture the essence of coordinated, evolutionary games as Assurance game properly describes such behaviors especially with reference to the biological communities [31, 32, 33, 39].

To construct a feasible model of the interactions for consequent tiger movements, we assume that the game on every iteration progresses by sharing of information between the players. Each interacting tiger obtains a finite number of information as an input from the contributing factor at each time step at a given state, and makes a transition to an unambiguously determined next state at the next time step. With this assumption, we construct a finite deterministic automaton to model the movement of the focal species [18, 19, 23]. We propose that the functional characterization of corridors, exchange information can be understood using a context free grammar, which is derived via the automata.

With the above discussions, we frame our research problem as: What is the finite deterministic automaton and the transition rule/ grammar that models the designing of wildlife corridors in the Indian Landscape?

Let $G(\Theta, \Sigma, \Pi)$ be a normal form, strategic game where $\forall i \in I = \{1,...,n\} \subset \aleph, n \geq 2$,

(i) $\Theta = \{\Theta_i\}$ is the set of interacting agents or players;

(ii) $\Sigma_i \neq \{\}$ is the set of strategies for the player $\Theta_i$. $\Sigma = \Sigma_1 \times ... \times \Sigma_n$ is the space of strategies, with $\sigma = (\sigma_1,...,\sigma_n) \in \Sigma$ being a strategy profile of the game $G$;

(iii) $\Pi_i : \Sigma \to \Re$ is the payoff function, which assigns to each strategy profile $\sigma$ a real number $\Pi_i(\sigma)$, the payoff earned by the player $\Theta_i$ when $\sigma$ is played in $G$. $\Pi = \Pi_1 \times ... \times \Pi_n$ is the space of payoff functions in the game.

Let the game $G$ be repeated in periods of discrete time $t \in \aleph$. Assume that the players are 'hardwired' to play only pure strategies in $G$. Thus each strategy set $\Sigma_i$ is a member of the standard basis for the strategy space $\Sigma$ where the $i^{th}$ coordinate is 1 and the rest are zeroes, and thus would correspond to a corner point of the simplex $\Lambda = \left\{ \hat{p} = (p_1, p_2,..., p_n)^T \in \Re : p_i \geq 0, i \in N, \sum_{i=1}^{n} p_i = 1 \right\}$, which is the simplex corresponding to $\Sigma$ [19].

Let an n-player Assurance game be represented by G(Θ,Σ,Π), where Θ = {Θ$_i$} is the set of players, with i ∈ ℑ={1,2,...,n} a finite index set and n ≥ 2. Σ = {Σ$_i$} where Σ$_i$ is the pure strategy set for each player Θ$_i$, with σ = {σ$_1$ ,σ$_2$ ,...,σ$_n$} where σ$_i$ ∈ Σ$_i$ for i ∈ ℑ is a pure strategy profile of the game and Π = {Π$_i$} , the set of pay-off functions Π$_i$: S →ℜ ∀ i ∈ ℑ where S is the set of strategy profiles, give the player's von Neumann-Morgenstern utility Π$_i$(σ) for every profile.

To obtain an understanding about the interactions between the parameters and tiger that leads to the designing of corridors in a landscape, we next study the Assurance game. The game is played between Assurance game players. For analyzing this n-player Assurance game, we consider the pay-off matrix adopted from [4]:

Table1: n-Player Assurance payoff matrix

| | Propotion of cooperators in the group | | | | | |
|---|---|---|---|---|---|---|
| | 100% | 80% | 60% | 40% | 20% | 0% |
| C | 20 | 14 | 8 | 1 | -8 | -15 |
| D | 6 | 6 | 6 | 6 | 6 | 6 |

The matrix shows pay-offs obtained by each player while playing against co-players who cooperate and interact. As could be observed from the payoff matrix, the scores from cooperative strategies for each player depend on the proportion of players who actually play C or play D in the entire population. It can be noticed that the pay-offs for the players playing C varies monotonically with the number of co-operators in the parameters. On the other hand, the pay-offs obtained by the player using the strategy of D remains constant irrespective of the number of players choosing to defect. This implies that the strategy selection of an Assurance game player depends on its expectations of the likely behavior of the similar players. If it expects that at least 80 % of its similar players in the parameters would play C, only then will it choose to play C. Else, it would change its strategy to D in order to obtain a higher pay-off.

A membership function (MF) is a curve that defines how each parameter in the input space is mapped to a membership value (or degree of membership) between 0 and 1. For the present modeling the input space is the total grid area which would considered for the landscape.

Chain code [12] is used to represent a boundary by a connected sequence of line. Typically this representation is based on 4 (or) 8 connectivity of the segments (as shown in figure1a, 1b). The direction of each segment is coded by using a numbering scheme.

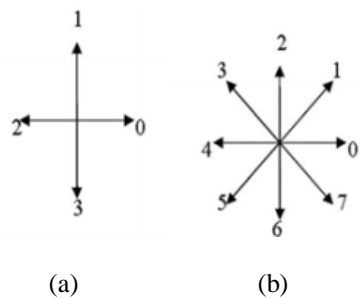

(a)          (b)

Fig 1. Neighbour directions of Chain Code

# 3. Modelling

For the purpose of the present work, we assume that the tiger habitat patches in India constitute the vertices and the collection of connections within this complex that connect any two of the habitats constitute the edges, comprising the focal landscape complex as a graph $\Gamma(V, E, \psi_\Gamma)$. The existence of an edge between any two vertices represents some ecological flux, such as animal movement, between the adjacent vertices.

To model the possible paths to serve as passages for tigers from one habitat patch to another habitat patch within any considered landscape complex, we first identify a set landscape factors or parameters, which may be natural or anthropogenic, and each of which may either constrain or support the passage of the tiger through the focal landscape matrix to various degrees, and hence become the major determinants in the structural connections becoming a corridor. For describing the present model, we consider five parameters *a, b, c, d* and *e*.

We assume that tigers in the landscape $(\Theta_1)$ and the set of above mentioned parameters of the landscape $(\Theta_2)$ constitute the two rational agents that play the Assurance game $G$ iterated over a number of generations. The players may use a number of strategies in the game in order to optimize their payoff. These payoffs are the costs incurred by the tiger population (called tiger henceforth in the paper) in using the landscape matrix for movement between habitats.

Next we code the different tiger habitats included in the focal landscape complex, by the following table:

Table 2. Coding for the tiger habitats in the complex

| S.No | Tiger habitat | Code |
|------|---------------|------|
| 1.   | Habitat 1     | 1    |
| 2.   | Habitat 2     | 2    |
| 3.   | Habitat 3     | 3    |
| 4.   | Habitat 4     | 4    |

In order to explain the model we create a random landscape image as shown in figure 2.

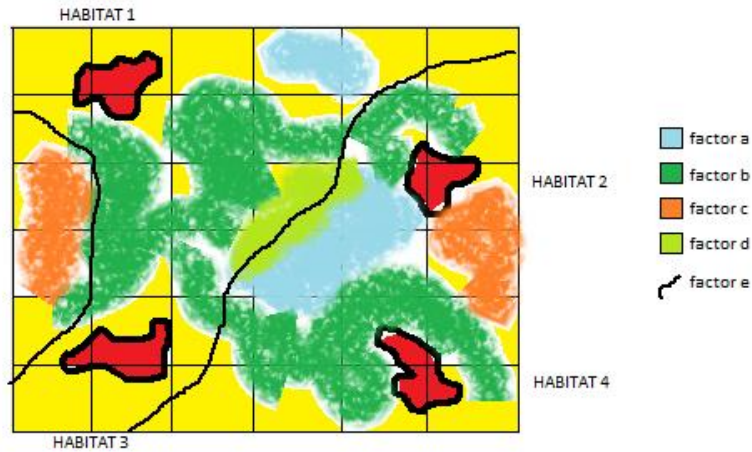

Fig 2. Sample landscape complex

Table 1 and the map in Fig. 2 lead to an adjacency matrix $A = [a_{ij}], i = 1,2,...,n; j = 1,2,...,n$, where n = 4 for tiger habitat patches, which can be seen in Table 3 and visualized through Figure 3.

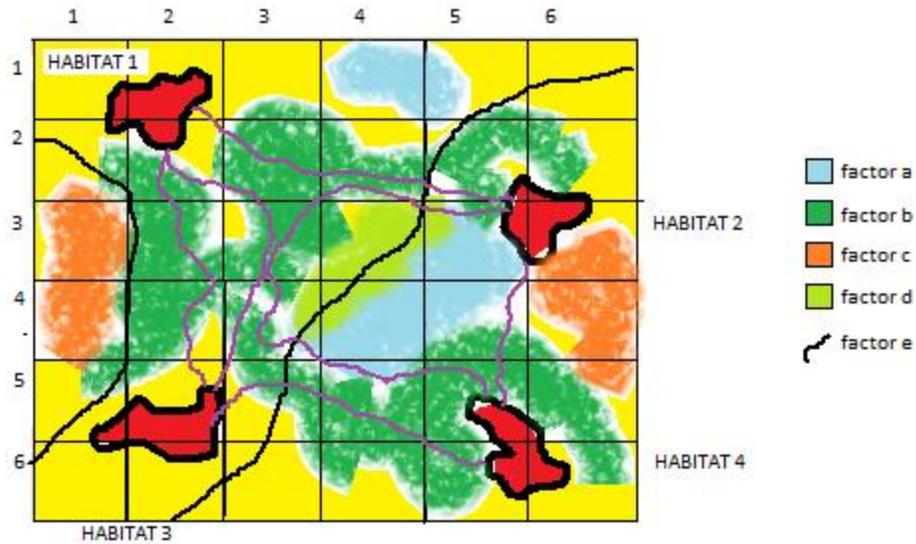

Fig 3. Sample landscape complex with connectivity.

Table 3. Adjacency matrix A=[$a_{ij}$] for tiger habitats in the sample landscape complex

|  | HABITAT 1 | HABITAT 2 | HABITAT 3 | HABITAT 4 |
|---|---|---|---|---|
| HABITAT 1 | 0 | 1 | 1 | 1 |
| HABITAT 2 | 1 | 0 | 1 | 1 |
| HABITAT 3 | 1 | 1 | 0 | 1 |
| HABITAT 4 | 1 | 1 | 1 | 0 |

From the obtained adjacency matrix we can check that there exists connectivity between every patch but the basic question lies in finding the most feasible connection of the entire set of connections that would facilitate the passage of tigers with minimal loss. In order to find out such path, we next compute the costs incurred on tigers in using the connections between different habitat patches in the given landscape complex.

$c : E \to \aleph$

$\ni e \mapsto c(e) = r \in \aleph, \forall e \in E, \aleph = \{0,1,...\}.$

In computing the cost matrix, we further create the Assurance game model using the contribution of each factor in the grid. Each factor of the landscape, due to its presence or absence contributes towards the cost matrix. For the present model, we consider 5 factors and categorize them as shown in Table 4.

Table 4. Factor categorization and score contribution

| Factor | Nature(Assumed) | Membership Contribution | | | | | | EXAMPLE |
|---|---|---|---|---|---|---|---|---|
| | | 1 | 0.8 | 0.6 | 0.4 | 0.2 | 0 | |
| a | Cooperative | 20 | 14 | 8 | 1 | -8 | -15 | WATER BODY |
| b | Cooperative | 20 | 14 | 8 | 1 | -8 | -15 | FOREST COVER |
| c | Defecting | 6 | 6 | 6 | 6 | 6 | 6 | AGRICULTURE LAND |
| d | Cooperative | 20 | 14 | 8 | 1 | -8 | -15 | PREY BASE |
| e | Defecting | 6 | 6 | 6 | 6 | 6 | 6 | HIGHWAYS |

For the purpose of scoring, we make few assumptions for our model, which can be perfectly calculated once worked on with the Remote Sensing and GIS data. The assumptions made are:

1. The area of each grid in the landscape is constant = $A$.
2. The area occupied by a factor $f$ in a grid $G_{ij}$ denotes the membership of the factor in the considered grid and is given by:
   $\mu_{f/Gij} = A_{f/Gij}/A$
3. The score of each parameter in a grid is based on its categorization and then application of bilinear interpolation between the values considered. For e.g. if
   $\mu_{a/G14} = .7$, then
   $\pi_{a/G14} = \frac{14(.7-.6)+8(.8-.7)}{(.7-.6)+(.8-.7)} = 11.$

Based on the above criteria of scoring, the various factors with respect to tiger using the membership of each factor in each grid and the strategy space of Assurance game the following cost matrix is obtained:

Table 5. Cost matrixes of the tiger for using existing corridors between different habitat patches in the complex

| | HABITAT 1 | HABITAT 2 | HABITAT 3 | HABITAT 4 |
|---|---|---|---|---|
| HABITAT 1 | 8 | S12 | S13 | S14 |
| HABITAT 2 | S21 | 8 | S23 | S24 |
| HABITAT 3 | S31 | S32 | 8 | S34 |
| HABITAT 4 | S41 | S42 | S43 | 8 |

For the present theoretical modelling we assume the following order of the scores, which can be correctly obtained using the presence, absence and abundance data of Remote Sensing and GIS:

$$S13 = S31 < S12 = S21 < S34 = S43 < S23 = S32 < S24 = S42 < S14 = S41$$

Using the above scores, we can rank the grids using the chain code algorithm which can be seen as:

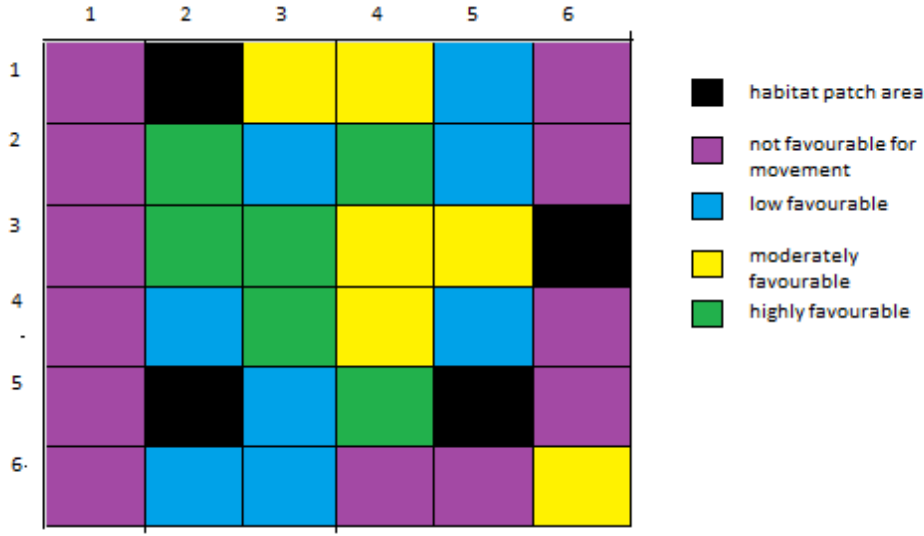

Fig. 4 Ranked grids using the cost matrix

Let the DFA that models the corridor designing and improving the landscape conditions for supporting movement of tigers be $\Delta(Q,\Sigma,q,\delta,h)$, where $Q$ is the set of states, $\Sigma$ is the alphabet, the set of input symbols or letters, $q \in Q$ is the initial state, $\delta$ is the transition function that prescribes the mapping of the automaton from one state to the next in time steps of $t \in N$, $h$ is the set of final states, the theorems in a Turing machine[31, 32, 36]. We list the objects comprising $\Delta$ in the following paragraphs:

Q comprises the following states, which represent the different states of grids that the tiger encounters while moving through it:

- Initial State(I)
- Not fovourable state(NFS)
- Fairly fovourable state(FFS)
- Moderately fovourable state(MFS)
- Fovourable state(FS)

The alphabet $\Sigma = \{a, b, c, d, e\}$ comprises the letters (inputs for the automata), which are the parameters present in the grid to play G.

I is the initial state, representing the initial state of a grid which appears as the tigers move out from the territorial region. The transition function $\delta$ is described by the following matrix:

Table 6. Transitions of Δ to various states

| Letter → / State ↓ | a | b | c | d | e |
|---|---|---|---|---|---|
| I | MFS | FS | NFS | FFS | NFS |
| NFS | FFS | FS | NFS | FFS | NFS |
| FFS | MFS | FS | NFS | MFS | NFS |
| MFS | FS | FS | FFS | FS | NFS |
| FS | FS | FS | FFS | FS | NFS |

There exist two states which may be included in the state of final states which are:

- NFS: Not Fovourable for movement of tigers and thus cannot be supported or converted to corridor due to massive interferences from inhibitory sources.
- FS: Fovourable State for movement, as it supports the movement of tigers through them with highest priority.

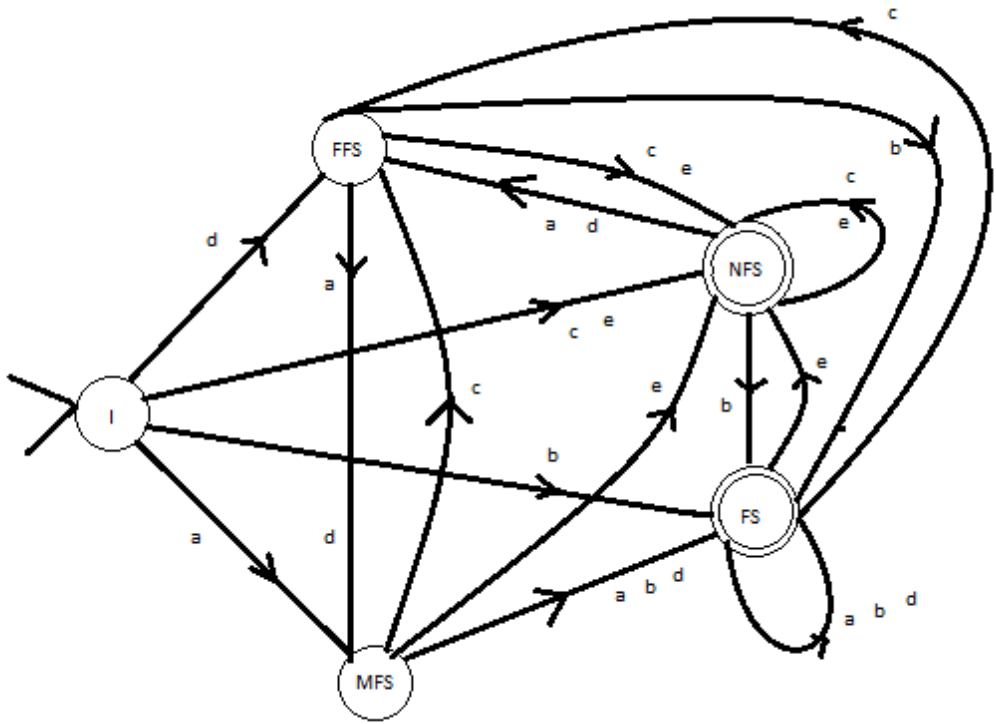

Fig. 5 DFA for grid state transitions to model wildlife corridors.

The above model of DFA would generate a standard context free grammar (CFG), which would, in essence, be determined by the transition function δ. Let the non-terminals in the DFA be denoted by Y. Y comprises the following states:

- I: Initial State (I)

- J: Not fovourable state (NFS)
- K: Fairly fovourable state (FFS)
- L: Moderately fovourable state (MFS)
- M: Fovourable state (FS)

Corresponding to such a set of non-terminal states, the context free grammar (CFG) couldbe written as

→ I→ cJ|aK|cL|Dm|eK

*J→ cJ|aK|cL|dM|eK

K→ cJ|aK|cL|dM|eK

L→ cJ|aK|cL|dM|eK

*M→ cJ|aK|cL|dM|eK

J, M is the final state of the automata, which, for the sake of identification, is prefixed by an asterisk sign.

## 4. Conclusion

The present work has been developed with objectives to (i) obtain a rule set to design a feasible tiger corridor network, connecting the habitat patches for the tiger in the landscape complex using a replicable computational procedure and (ii) identify the most important habitat patches, along with their underlying community structure so as to focus efforts towards conserving them.

In this paper, we have used Deterministic Finite Automaton to obtain a grammar that could serve as a model framework for a real-world tiger corridor designing in the Indian landscape.

A limitation in the modelling described in the paper is that the corridor designing is based entirely on the structural definition of connectivity, and thus does not take into account some critically vital landscape features such as the biotic factors of availability of prey base and water, in computing the cost matrix. The work is, by choice, kept rudimentary so as to provide a basic computational framework for perceiving a viable structural corridor network design in the focal landscape complex for tigers. We may justify this absence of path redundancy consideration due to two reasons: first, our priority in the paper was to focus on network efficiency over redundancy, and second, the work focuses on estimation of optimal strategies for connecting the CTHs, rather than inclusion of alternative paths [55]. We are aware that such a simplification is more often not in consonance with the real-world corridor scenario. We however hope that our present effort would make available a computational template for tiger corridor designing, which could certainly be improved upon by incorporating field data from realistic considerations.